\documentclass[aps,prl]{revtex4}
\usepackage{epsf}
\usepackage{bm}

\begin{document}

\title{CONFIGURATION SPACE METHOD FOR CALCULATING\\
BINDING ENERGIES OF EXCITON COMPLEXES IN QUASI-1D/2D SEMICONDUCTORS}

\author{\footnotesize I.V. BONDAREV}

\address{Department of Mathematics and Physics\\
North Carolina Central University\\
1801 Fayetteville Str, Durham, North Carolina 27707, USA\\ibondarev@nccu.edu}

\begin{abstract}
A configuration space method is developed for binding energy calculations of the lowest energy exciton complexes (trion, biexciton) in spatially confined quasi-1D semiconductor nanostructures such as  nanowires and nanotubes. Quite generally, trions are shown to have greater binding energy in strongly confined structures with small reduced electron-hole masses. Biexcitons have greater binding energy in less confined structures with large reduced electron-hole masses. This results in a universal crossover behavior, whereby trions become less stable than biexcitons as the transverse size of the quasi-1D nano\-structure increases. The method is also capable of evaluating binding energies for electron-hole complexes in quasi-2D semiconductors such as coupled quantum wells and bilayer van der Walls bound heterostructures with advanced optoelectronic properties.
\end{abstract}

\maketitle

\section{Introduction}\label{sec1}

Optical properties of low-dimensional semiconductor nanostructures originate from excitons (Coulomb-bound electron-hole pairs) and exciton complexes such as biexcitons (coupled states of two excitons) and trions (charged excitons). These have pronounced binding energies in nanostructures due to the quantum confinement effect.\cite{HaugKoch,Cardona,Apphysrev} The advantage of optoelectronic device applications with low-dimensional semiconductor nanostructures lies in the ability to tune their properties in a controllable way. Optical properties of semiconducting carbon nanotubes (CNs), in particular, are largely determined by excitons,\cite{Dresselhaus07,Louie09} and can be tuned by electrostatic doping,\cite{Steiner09,Spataru10,Mueller10,Kato} or by means of the quantum confined Stark effect.\cite{Bondarev09PRB,Bondarev12PRB,Bondarev14PRBbec,Pedersen15} Carbon nanotubes are graphene sheets rolled-up into cylinders of one to a few nanometers in diameter and up to hundreds of microns in length, which can be both metals and semiconductors depending on their diameters and chirality.\cite{Saito,Dresselhaus} Over the past decade, optical nanomaterials research has uncovered intriguing optical attributes of their physical properties, lending themselves to a variety of new optoelectronic device applications.\cite{Papanikolas,Imamoglu,Avouris08,McEuen09,Hertel10,Bondarev10jctn,Strano11,BondarevNova11,Kono12,Baughman13,ChemPhysSI,BondarevOE15}

Formation of biexcitons and trions, though not detectable in bulk materials at room temperature, play a significant role in quantum confined systems of reduced dimensionality such as quantum wells,\cite{Birkedal,Singh,Thilagam,Lozovik,Bracker} nanowires,\cite{Forchel,Crottini,Sidor,Gonzales,Schuetz} nanotubes,\cite{Pedersen03,Pedersen05,Kammerlander07,Ronnow10,Ronnow11,Matsunaga11,Santos11,Bondarev11PRB,Bondarev14PRB,Watanabe12,Ronnow12,Colombier12,Yuma13} and quantum dots.\cite{Woggon,JonFbiexc,JonFtrion} Biexciton and trion excitations open up routes for controllable nonlinear optics and spinoptronics applications, respectively. The trion, in particular, has both net charge and spin, and therefore can be controlled by electrical gates while being used for optical spin manipulation, or to investigate correlated carrier dynamics in low-dimensional materials.

\begin{figure}[t]
\epsfxsize=17.5cm\centering{\epsfbox{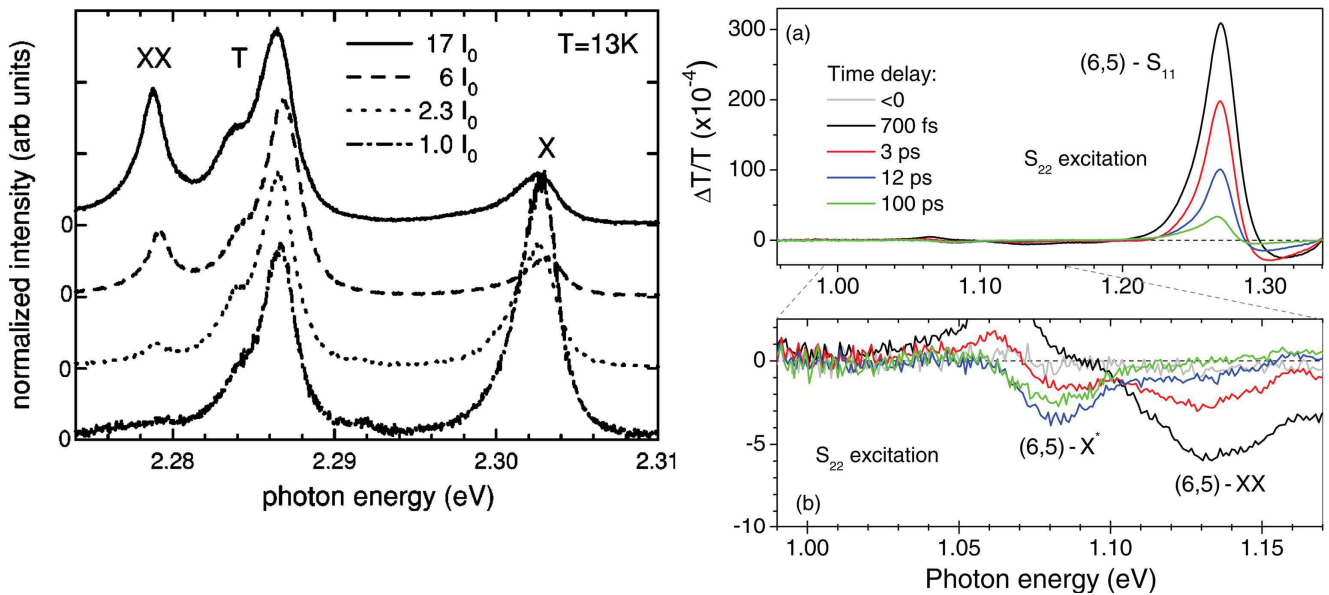}}\caption{Left panel: An example of the time-integrated photoluminescence spectra of a single CdSe quantum dot at multiple excitation intensities ($I_0\!=\!250~\mu$W). The spectra are normalized to the peak at 2.287~eV and are displaced vertically for clarity. Exciton (X) and trion (T) emission is present at the lowest excitation intensities. The biexciton (XX) emission peak shows up as the excitation intensity increases at the photon energy \emph{lower} than that of the trion peak, indicating the \emph{greater} biexciton binding energy than that of the trion. Reprinted with permission from Ref.[51]. Right panel: (a) Differential transmission spectra obtained for different probe laser time-delays after the excitation of the second bright exciton transition ($S_{22}$ excitation) of the semiconducting (6,5) CN by the pump laser with fluence $6.3\times10^{13}$ photons/pulse/cm$^2$. Induced transmission (IT) at 1.26~eV is due to the fast second-to-first bright exciton ($S_{22}\!\rightarrow\!S_{11}$) relaxation. (b)~Magnification of the spectral region below the energy of the first bright exciton excitation energy ($S_{11}$) in the (6,5) nanotube. Induced absorption (IA) at 1.08~eV that appears at longer time-delays is attributed to trion ($X^*$) formation. The short time-delay IA (black line) is due to biexciton ($XX$) formation at the photon energy 1.13~eV, which is \emph{greater} than the trion IA energy, indicating the \emph{lower} biexciton binding energy than that of the trion. Reprinted with permission from Ref.[50].}\label{fig0}
\end{figure}

For conventional semiconductor quantum wells, wires, and dots, the binding energies of negatively or positively charged trions are known to be typically lower than those of biexcitons in the same nanostructure, although the specific trion to biexciton binding energy ratios are strongly sample fabrication dependent.\cite{Lozovik,Forchel,Sidor,Woggon} First experimental evidence for the trion formation in carbon nanotubes was reported by Matsunaga et al.\cite{Matsunaga11} and by Santos et al.\cite{Santos11} on $p$-doped (7,5) and undoped (6,5) CNs, respectively. Theoretically, R{\o}nnow et al.\cite{Ronnow10} have predicted that the lowest energy trion states in all semiconducting CNs with diameters of the order of or less than 1~nm should be stable at room temperature. They have later developed the fractional dimension approach to simulate binding energies of trions and biexcitons in quasi-1D/2D semiconductors, including nanotubes as a particular case.\cite{Ronnow11,Ronnow12} Binding energies of $63$~meV and $92$~meV are reported for the lowest energy trions\cite{Ronnow11} and biexcitons,\cite{Ronnow12} respectively, in the (7,5) nanotube. However, the recent nonlinear optics experiments were able to resolve both trions and biexcitons in the same CN sample,\cite{Colombier12,Yuma13} to report on the opposite tendency where the binding energy of the trion \emph{exceeds} that of the biexciton rather significantly in small diameter ($\lesssim\!1$~nm) CNs. Figure~\ref{fig0} shows typical experimental data for conventional low-dimension semiconductors (left panel) and small diameter semicondicting CNs (right panel). In the left panel, the biexciton resonance is seen to appear at lower photon energy than the trion one, in contrast with the right panel where the biexciton resonance manifests itself at greater photon energy than the trion resonance does. This clearly indicates greater trion binding energies than those of biexcitons in small diameter semiconducting CNs as opposed to conventional low-dimension semiconductors.

\begin{figure}[t]
\epsfxsize=10.0cm\centering{\epsfbox{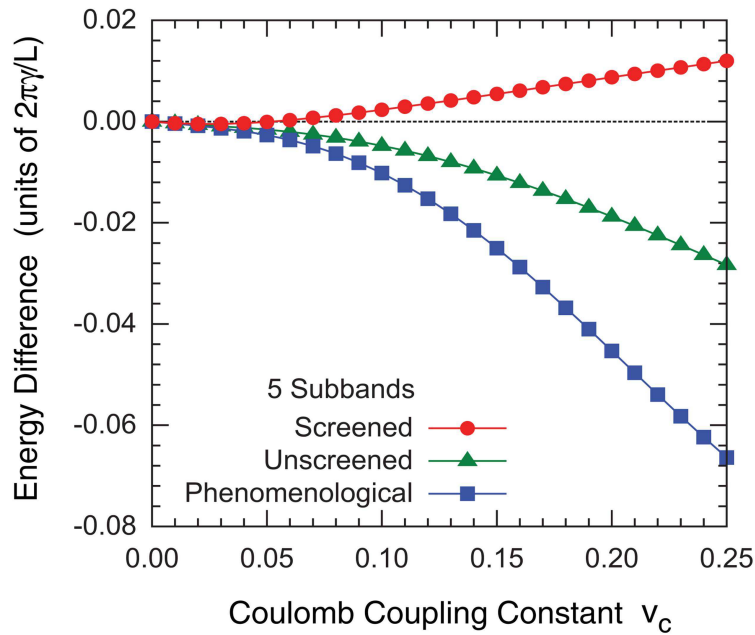}}\caption{Differences between the trion and biexciton binding energies (in units of $2\pi\gamma/L$, where $L$ is the CN circumference and $\gamma\!=\!6.46$~eV$\,$\AA) as functions of the Coulomb coupling parameter $v_c$, which are obtained within the phenomenological, unscreened and screened models. The parameter $v_c$ is approximately equal to the ratio of the Coulomb energy to the CN band gap (typically $\lesssim\!0.25$). Reprinted with permission from Ref.[47].}\label{fig00}
\end{figure}

More specifically, Colombier et al.\cite{Colombier12} reported on the observation of the binding energies $150$~meV and $106$~meV for the trion and biexciton, respectively, in the (9,7) CN. Yuma et al.\cite{Yuma13} reported even greater binding energies of $190$~meV for the trion versus $130$~meV for the biexciton in the smaller diameter (6,5) CN. (Their spectra are reproduced in Fig.~\ref{fig0}, right panel.) In both cases, the trion-to-biexciton binding energy ratio is greater than unity, decreasing as the CN diameter increases [1.46 for the 0.75~nm diameter (6,5) CN versus 1.42 for the 1.09~nm diameter (9,7) CN]. Trion binding energies greater than those of biexcitons are theoretically reported by Watanabe and Asano,\cite{Watanabe12} due to the energy band nonparabolicity and the Coulomb screening effect that reduces the biexciton binding energy more than that of the trion. Watanabe and Asano have extended the first order ($\mathbf{k}\cdot\mathbf{p}$)-perturbation series expansion model originally developed by Ando for excitons (see Ref.\cite{Ando2005} for review) to the case of electron-hole complexes such as trions and biexcitons. Figure~\ref{fig00} compares the differences between the trion and biexciton binding energies delivered by "phenomenological" and "unscreened" models termed as such to refer to the cases where the energy band nonparabolicity, electron-hole complex form-factors, self-energies and the screening effect are all neglected, and where all of them but screening are taken into account, respectively, with the difference given by the "screened" model. The latter is the Watanabe--Asano model which includes \emph{all} of the factors mentioned within the first order ($\mathbf{k}\cdot\mathbf{p}$)-perturbation theory. One can see that the "screened" model does predict greater trion binding energies than those of biexcitons as opposed to the phenomenological and unscreened models. However, the most the trion binding energy can exceed that of the biexciton within this model is $0.012\,(2\pi\gamma/L)$ equal to $21$ and $14$~meV for the (6,5) and (9,7) CNs, respectively, which is obviously not enough to explain the experimental observations.


This article reviews the capabilities of the configuration space (Landau-Herring) method for the binding energy calculations of the lowest energy exciton complexes in quasi-1D/2D semiconductors. The approach was originally pioneered by Landau,\cite{LandauQM} Gor'kov and Pitaevski,\cite{Pitaevski63} Holstein and Herring\cite{Herring} in the studies of molecular binding and magnetism. The method was recently shown to be especially advantageous in the case of quasi-1D semiconductors,\cite{Bondarev11PRB,Bondarev14PRB} allowing for easily tractable, complete analytical solutions to reveal universal asymptotic relations between the binding energy of the exciton complex of interest and the binding energy of the exciton in the same nanostructure. The Landau-Herring method of the complex bound state binding energy calculation is different from commonly used quantum mechanical approaches reviewed above. These either use advanced simulation techniques to solve the coordinate-space Schr\"{o}dinger equation numerically,\cite{Pedersen05,Kammerlander07,Ronnow10,Ronnow11,Ronnow12} or convert it into the reciprocal (momentum) space to follow up with the ($\mathbf{k}\cdot\mathbf{p}$)-perturbation series expansion calculations.\cite{Watanabe12,Ando2005} Obviously, this latter one, in particular, requires for perturbations to be small. If they are not, then the method brings up an underestimated binding energy value, especially for molecular complexes such as biexciton and trion where the kinematics of complex formation depends largely on the asymptotic behavior of the wave functions of the constituents. This is likely the cause for Watanabe--Asano theory of excitonic complexes\cite{Watanabe12} to significantly underestimate the measurements by Colombier et al.\cite{Colombier12} and Yuma et al.\cite{Yuma13} on semiconducting CNs.


\begin{figure}[t]
\epsfxsize=11.5cm\centering{\epsfbox{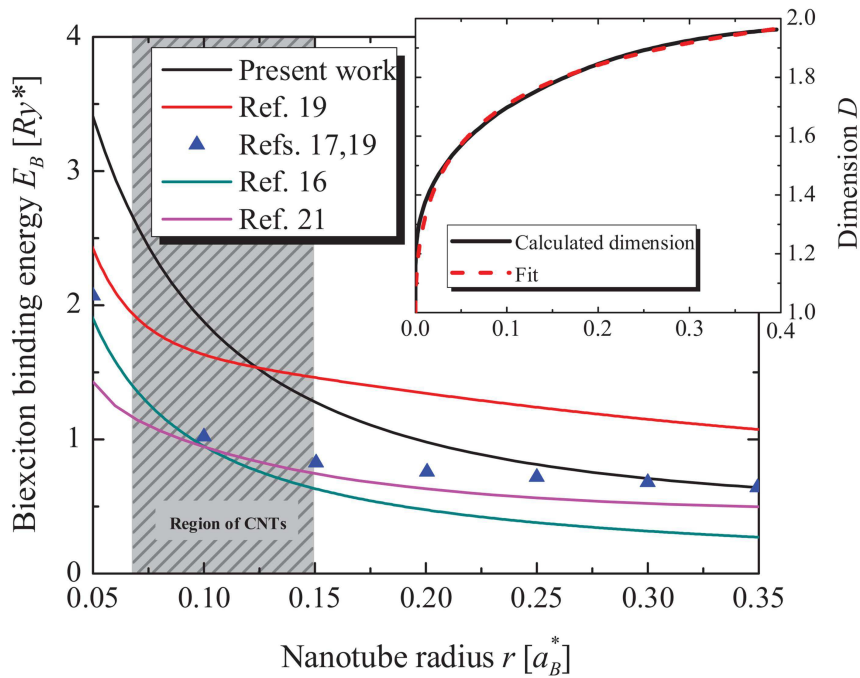}}\caption{Comparison of the biexciton binding energy given by the fractional dimension model of Ref.[48] to other coordinate-space formulated models ("phenomenological" in terms of Fig.~\ref{fig00}) and the configuration space model. Dimensionless units are used (defined in Sec.~2). Inset shows the relation between the effective CN radius and the dimension $D$ calculated according to Ref.[41]. Reference number correspondence to this article is as follows: "Present work"$\rightarrow\,$Ref.[48], "Refs.17,19"$\rightarrow\,$Ref.[40], "Ref.16"$\rightarrow\,$Ref.[39] (all of these are the phenomenological models in terms of Fig.~\ref{fig00}), "Ref.21"$\,\rightarrow\,$Ref.[45] (the configuration space model used in this article). Reprinted with permission from Ref.[48].}\label{fig000}
\end{figure}

The Landau-Herring configuration space approach does not have this shortcoming. It works in the \emph{configuration space} of the \emph{two} relative electron-hole motion coordinates of the \emph{two} non-interacting quasi-1D excitons that are modeled by the effective one-dimensional cusp-type Coulomb potential as proposed by Ogawa and Takagahara for 1D semiconductors.\cite{Ogawa91} Since the configuration space is different from the ordinary coordinate (or its reciprocal momentum) space, the approach does not belong to any of the models summarized in Fig.~\ref{fig00}. In this approach, the biexciton or trion bound state forms due to the exchange under-barrier tunneling between the equivalent configurations of the electron-hole system in the configuration space. The strength of the binding is controlled by the exchange tunneling rate. The corresponding binding energy is given by the tunnel exchange integral determined through an appropriate variational procedure. As any variational approach, the method gives an upper bound for the \emph{ground} state binding energy of the exciton complex of interest. As an example, Fig.~\ref{fig000} compares the biexciton binding energies calculated within several different models, including those coordinate-space formulated that are referred to as phenomenological in Fig.~\ref{fig00}, as well as the configuration space model. It is quite remarkable that with obvious overall correspondence to the other methods as seen in Fig.~\ref{fig000}, the Landau-Herring configuration space approach is the only to have been able consistently explain the experimental observations discussed above and shown in Fig.~\ref{fig0}, both for conventional low-dimension semiconductors and for semiconducting CNs. Whether the trion or biexciton is more stable (has greater binding energy) in a particular quasi-1D system turns out to depend on the reduced electron-hole mass and on the characteristic transverse size of the system.\cite{Bondarev14PRB} Trions are generally more stable than biexcitons in strongly confined quasi-1D structures with small reduced electron-hole masses, while biexcitons are more stable than trions in less confined quasi-1D structures with large reduced electron-hole masses. As such, a crossover behavior is predicted,\cite{Bondarev14PRB} whereby trions get less stable than biexcitons as the transverse size of the quasi-1D nanostructure increases --- quite a general effect which could likely be observed through comparative measurements on semiconducting CNs of increasing diameter. The method captures the essential kinematics of exciton complex formation, thus helping understand in simple terms the general physical principles that underlie experimental observations on biexcitons and trions in a variety of quasi-1D semiconductor nano\-structures. For semiconducting CNs with diameters $\lesssim1\!$~nm, the model predicts the trion binding energy greater than that of the biexciton by a factor $\sim\!1.4$ that decreases with the CN diameter increase, in reasonable agreement with the measurements by Colombier et al.\cite{Colombier12} and Yuma et al.\cite{Yuma13}

The article is structured as follows. Section 2 formulates the general Hamiltonian for the biexciton complex of two electrons and two holes in quasi-1D semiconductor. Carbon nanotubes of varying diameter are used as a model example for definiteness. The theory and conclusions are valid for any quasi-1D semiconductor system in general. The exchange integral and the binding energy of the biexciton complex are derived and analyzed. Section 3 further develops the theory to include the trion case. In Section 4, the trion binding energy derived is compared to the biexciton binding energy for semiconducting quasi-1D nanostructures of varying transverse size and reduced exciton effective mass. Section 5 generalizes the method to include trion and biexciton complexes formed by indirect excitons in layered quasi-2D semiconductor structures such as coupled quantum wells (CQWs) and bilayer self-assembled transition metal dichalchogenide heterostructures. Section 6 summarizes and concludes the article.

\section{Biexciton in quasi-1D}\label{sec2}

\begin{figure}[t]
\epsfxsize=11.0cm\centering{\epsfbox{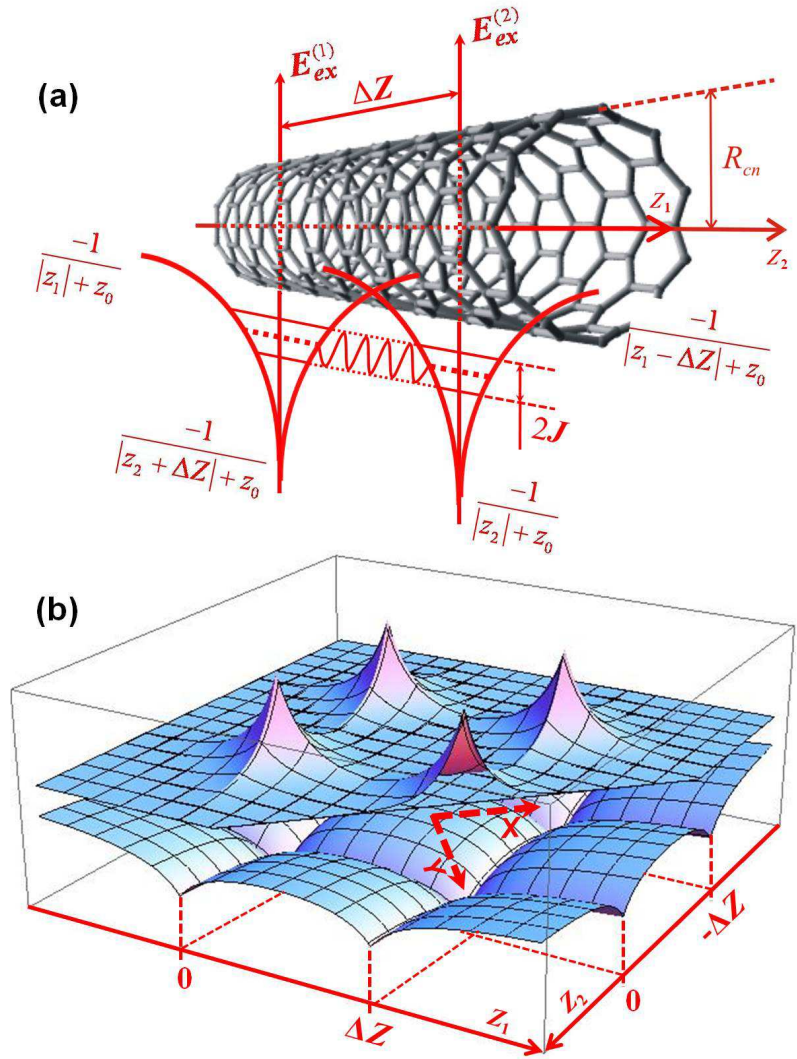}}\caption{(a)~Schematic of the exchange coupling of two ground-state 1D excitons to form a~biexcitonic~state (arb.~units). Two collinear axes, $z_1$ and $z_2$, representing independent relative electron-hole motions in the 1st and 2nd exciton, have their origins shifted by $\Delta Z$, the inter-exciton center-of-mass separation. (b)~The coupling occurs in the configuration space of the two independent longitudinal relative electron-hole motion coordinates, $z_1$ and $z_2$, of each of the excitons, due to the tunneling of the system through the potential barriers formed by the two single-exciton cusp-type potentials [bottom, also in (a)], between equivalent states represented by the isolated two-exciton wave functions shown on the top.}\label{fig1}
\end{figure}

The problem is initially formulated for two interacting ground-state 1D excitons in a semiconducting CN. The CN is taken as a model for definiteness. The theory and conclusions are valid for any quasi-1D semiconductor system in general. The excitons are modeled by effective one-dimensional cusp-type Coulomb potentials, shown in Fig.~\ref{fig1}~(a), as proposed by Ogawa and Takagahara for 1D semiconductors.\cite{Ogawa91} The intra-exciton motion can be legitimately treated as being much faster than the inter-exciton center-of-mass relative motion since the exciton itself is normally more stable than any of its compound complexes. Therefore, the adiabatic approximation can be employed to simplify the formulation of the problem. With this in mind, using the cylindrical coordinate system [\emph{z}-axis along the CN as in Fig.~\ref{fig1}~(a)] and separating out circumferential and longitudinal degrees of freedom for each of the excitons by transforming their longitudinal motion into their respective center-of-mass coordinates,\cite{Bondarev09PRB,Ogawa91} one arrives at the Hamiltonian of the form\cite{Bondarev14PRB}
\begin{eqnarray}
\hat{H}(z_1,z_2,\Delta Z)=-\frac{\partial^2}{\partial\,\!z_{1}^2}-\frac{\partial^2}{\partial\,\!z_{2}^2}\hskip2cm\label{biexcham}\\
-\frac{1}{|z_{1}|\!+\!z_0}-\!\frac{1}{|z_{1}\!-\!\Delta Z|\!+\!z_0}-\!\frac{1}{|z_{2}|\!+\!z_0}-\!\frac{1}{|z_{2}\!+\!\Delta Z|\!+\!z_0}\hskip0.3cm\nonumber\\
-\frac{2}{|(\sigma z_1+z_2)/\lambda+\Delta Z|\!+\!z_0}-\frac{2}{|(z_1+\sigma z_2)/\lambda-\Delta Z|\!+\!z_0}\nonumber\\
+\frac{2}{|\sigma(z_1-z_2)/\lambda+\Delta Z|\!+\!z_0}+\frac{2}{|(z_1-z_2)/\lambda-\Delta Z|\!+\!z_0}\;.\nonumber
\end{eqnarray}
Here, $z_{1,2}\!=\!z_{e1,2}-z_{h1,2}$ are the relative electron-hole motion coordinates of the two 1D excitons separated by the center-of-mass-to-center-of-mass distance $\Delta Z\!=\!Z_2-Z_1$, $z_0$ is the cut-off parameter of the effective (cusp-type) longitudinal electron-hole Coulomb potential, $\sigma\!=\!m_e/m_h$, $\lambda\!=\!1+\sigma$ with $m_e$ ($m_h$) representing the electron (hole) effective mass. The "atomic units"\space are used,\cite{LandauQM,Pitaevski63,Herring} whereby distance and energy are measured in units of the exciton Bohr radius $a^\ast_B\!=\!0.529\,\mbox{\AA}\,\varepsilon/\mu$ and the Rydberg energy $Ry^\ast\!=\hbar^2/(2\mu\,m_0a_B^{\ast2})\!=\!13.6\,\mbox{eV}\,\mu/\varepsilon^2$, respectively, $\mu\!=\!m_e/(\lambda\,m_0)$ is the exciton reduced mass (in units of the free electron mass $m_0$) and $\varepsilon$ is the static dielectric constant of the electron-hole Coulomb potential.


The first two lines in Eq.~(\ref{biexcham}) represent two non-interacting 1D excitons. Their individual potentials are symmetrized to account for the presence of the neighbor a distance~$\Delta Z$ away, as seen from the $z_1$- and $z_2$-coordinate systems treated independently [Fig.~\ref{fig1}~(a)]. The last two lines are the inter-exciton exchange Coulomb interactions --- electron-hole (line next to last) and hole-hole + electron-electron (last line), respectively. The binding energy $E_{X\!X}$ of the biexciton is given by the difference $E_g\!-2E_X$, where $E_g$ is the lowest eigenvalue of the Hamiltonian~(\ref{biexcham}) and $E_X\!=-Ry^\ast/\nu_0^2$ is the single-exciton binding energy with $\nu_0$ being the lowest-bound-state quantum number of the 1D exciton.\cite{Ogawa91} Negative $E_{X\!X}$ indicates that the biexciton is stable with respect to the dissociation into two isolated excitons. The strong transverse confinement in reduced dimensionality semiconductors is known to result in the mass reversal effect,\cite{HaugKoch,Cardona} whereby the bulk heavy hole state, that forming the \emph{lowest} excitation energy exciton, acquires a longitudinal mass comparable to the bulk \emph{light} hole mass ($\approx\!m_e$). Therefore, $m_h\!\approx\!m_e$ in our case of interest here, which is also true for graphitic systems such as CNs, in particular,\cite{Jorio05} and so $\sigma\!=\!1$ is assumed in Eq.~(\ref{biexcham}) in what follows with no substantial loss of generality.

The Hamiltonian (\ref{biexcham}) is effectively two dimensional in the configuration space of the two \emph{independent} relative motion coordinates, $z_1$ and $z_2$. Figure~\ref{fig1}~(b), bottom, shows schematically the potential energy surface of the two closely spaced non-interacting 1D excitons [second line of Eq.~(\ref{biexcham})] in the $(z_1,z_2)$ space. The surface has four symmetrical minima representing isolated two-exciton states shown in Fig.~\ref{fig1}~(b), top. These minima are separated by the potential barriers responsible for the tunnel exchange coupling between the two-exciton states in the configuration space. The coordinate transformation
\begin{equation}
x=\frac{z_1-z_2-\Delta Z}{\sqrt{2}}\,,\hskip1cm y=\frac{z_1+z_2}{\sqrt{2}}
\label{transformation}
\end{equation}
places the origin of the new coordinate system into the intersection of the two tunnel channels between the respective potential minima [Fig.~\ref{fig1}~(b)], whereby the exchange splitting formula of Refs.\cite{LandauQM,Pitaevski63,Herring} takes the form
\begin{equation}
E_{g,u}(\Delta Z)-2E_X=\mp J(\Delta Z).
\label{Egu}
\end{equation}
Here $E_{g,u}(\Delta Z)$ are the ground-state and excited-state energies, eigenvalues of the Hamiltonian~(\ref{biexcham}), of the two coupled excitons as functions of their center-of-mass-to-center-of-mass separation, and $J(\Delta Z)$ is the tunnel exchange coupling integral responsible for the bound state formation of two excitons. For biexciton, this takes the form
\begin{equation}
J_{X\!X}(\Delta Z)=\frac{2}{3!}\int_{\!-\Delta Z/\!\sqrt{2}}^{\Delta Z/\!\sqrt{2}}\!dy\left|\psi_{X\!X}(x,y)\frac{\partial\psi_{X\!X}(x,y)}{\partial x}\right|_{x=0},
\label{JXX}
\end{equation}
where $\psi_{X\!X}(x,y)$ is the solution to the Schr\"{o}dinger equation with the Hamiltonian~(\ref{biexcham}) transformed to the $(x,y)$ coordinates. The factor $2/3!$ comes from the fact that there are two equivalent tunnel channels in the biexciton problem, mixing three equivalent indistinguishable two-exciton states in the configuration space --- one state is given by the two minima on the $x$-axis and two more are represented by each of the minima on the $y$-axis [cf. Fig.~\ref{fig1}~(a) and Fig.~\ref{fig1}~(b)].

The function $\psi_{X\!X}(x,y)$ in Eq.~(\ref{JXX}) is sought in the form
\begin{equation}
\psi_{X\!X}(x,y)=\psi_0(x,y)\exp[-S_{X\!X}(x,y)]\,,
\label{psiXXxy}
\end{equation}
where
\begin{equation}
\psi_0(x,y)=\frac{1}{\nu_0}\exp\!\left[-\frac{1}{\nu_0}\left(|z_1(x,y,\Delta Z)|+|z_2(x,y,\Delta Z)|\right)\right]
\label{psi0xy}
\end{equation}
is the product of two single-exciton wave functions (ground state) representing the isolated two-exciton state centered at the minimum $z_1\!=\!z_2\!=\!0$ (or $x\!=\!-\Delta Z/\sqrt{2}$, $y\!=\!0$) of the configuration space potential [Fig.~\ref{fig1}~(b)]. This is the approximate solution to the Shr\"{o}dinger equation with the Hamiltonin given by the first two lines in Eq.~(\ref{biexcham}), where the cut-off parameter $z_0$ is neglected.\cite{Ogawa91} This approximation greatly simplifies problem solving, while still remaining adequate as only the long-distance tail of $\psi_0$ is important for the tunnel exchange coupling. The function $S_{X\!X}(x,y)$, on the other hand, is a slowly varying function to account for the major deviation of $\psi_{X\!X}$ from $\psi_0$ in its "tail area" due to the tunnel exchange coupling to another equivalent isolated two-exciton state centered at $z_1\!=\Delta Z$, $z_2\!=\!-\Delta Z$ (or $x\!=\!\Delta Z/\sqrt{2}$, $y\!=\!0$).

Substituting Eq.~(\ref{psiXXxy}) into the Schr\"{o}dinger equation with the Hamiltonian (\ref{biexcham}) pre-transformed to the $(x,y)$ coordinates, one obtains in the region of interest ($z_0$ dropped for the reason above)
\[
\frac{\partial S_{X\!X}}{\partial x}=\nu_0\left(\frac{1}{x+3\Delta Z/\sqrt{2}}-\frac{1}{x-\Delta Z/\sqrt{2}}+\frac{1}{y-\sqrt{2}\Delta Z}-\frac{1}{y+\sqrt{2}\Delta Z}\right),
\]
up to negligible terms of the order of the inter-exciton~van der Waals energy and up to the second order derivatives of~$S_{X\!X}$. This equation is to be solved with the boundary condition $S_{X\!X}(-\Delta Z/\sqrt{2},y)\!=\!0$ originating from the natural requirement $\psi_{_{X\!X}}(-\Delta Z/\sqrt{2},y)\!=\!\psi_0(-\Delta Z/\sqrt{2},y)$, to result in
\begin{equation}
S_{X\!X}(x,y)=\nu_0\!\left(\!\ln\!\left|\frac{x\!+\!3\Delta Z/\!\sqrt{2}}{x-\Delta Z/\!\sqrt{2}}\right|+\frac{2\sqrt{2}\Delta Z(x\!+\!\Delta Z/\!\sqrt{2})}{y^2-2\Delta Z^2}\!\right)\!.
\label{sXXxy}
\end{equation}
After plugging this into Eq.~(\ref{psiXXxy}) one can calculate the tunnel exchange coupling integral~(\ref{JXX}). Retaining only the leading term of the integral series expansion in powers of $\nu_0$ subject to $\Delta Z>1$, one obtains
\begin{equation}
J_{X\!X}(\Delta Z)=\frac{2}{3\nu_0^3}\left(\frac{e}{3}\right)^{2\nu_0}\!\Delta Z\,e^{-2\Delta Z/\nu_0}.
\label{JXXfin}
\end{equation}

The ground state energy $E_g(\Delta Z)$ of the two coupled 1D excitons in Eq.~(\ref{Egu}) is now seen to go through the negative minimum (biexcitonic state) as $\Delta Z$ increases. The minimum occurs at $\Delta Z_0=\nu_0/2$, whereby the biexciton binding energy takes the form
\[
E_{X\!X}=-J_{X\!X}(\nu_0/2)=-\frac{1}{9\nu_0^2}\left(\frac{e}{3}\right)^{2\nu_0-1}
\]
in atomic units. Expressing $\nu_0$ in terms of $E_X$, one obtains in absolute units the equation as follows
\begin{equation}
E_{X\!X}=-\frac{1}{9}\;|E_X|\left(\frac{e}{3}\right)^{2\sqrt{Ry^\ast/|E_X|}\,-\,1}\!\!\!.
\label{Exx}
\end{equation}


\begin{figure}[t]
\epsfxsize=11.0cm\centering{\epsfbox{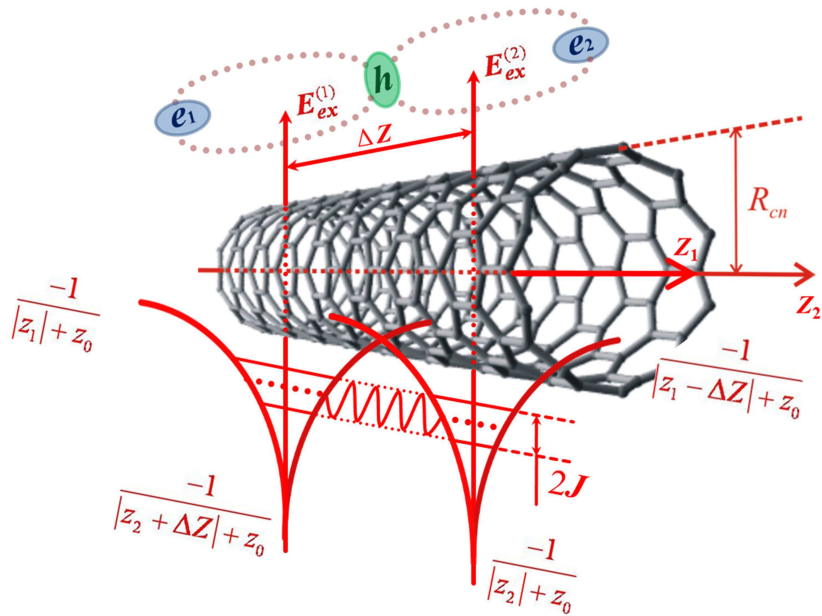}}\caption{Schematic of the two ground-state 1D excitons sharing the same hole to form a negative trion state (arb.~units). Two collinear axes, $z_1$ and $z_2$, representing independent relative electron-hole motions in the 1st and 2nd exciton, have their origins shifted by $\Delta Z$, the inter-exciton center-of-mass distance.}\label{fig2}
\end{figure}

\section{Trion in quasi-1D}\label{sec3}

The trion binding energy can be found in the same way using a modification of the Hamiltonian (\ref{biexcham}), in which two same-sign particles share the third particle of an opposite sign to form the two equivalent 1D excitons as Fig.~\ref{fig2} shows for the negative trion complex consisting of the hole shared by the two electrons. The Hamiltonian modified to reflect this fact has the first two lines exactly the same as in Eq.~(\ref{biexcham}), no line next to last, and one of the two terms in the last line --- either the first or the second one for the positive (with $z_{1,2}\!=\!z_{e}-z_{h1,2}$) and negative (with $z_{1,2}\!=\!z_{e1,2}-z_h$) trion, respectively. Obviously, due to the additional mass factor $\sigma$ (typically less than one for bulk semiconductors) in the hole-hole interaction term in the last line, the positive trion might be expected to have a greater binding energy in this model, in agreement with the results reported earlier.\cite{Sidor,Ronnow10} However, as was already mentioned in Sec.~\ref{sec2}, the mass reversal effect in \emph{strongly} confined reduced dimensionality semiconductors is to result in $\sigma\!=\!1$ in the trion Hamiltonian. The positive-negative trion binding energy difference disappears then. The negative trion case illustrated in Fig.~\ref{fig2}, is addressed below.

Just like in the case of the biexciton, the treatment of the trion problem starts with the coordinate transformation~(\ref{transformation}) to bring the trion Hamiltonian from the original (configuration space) coordinate system $(z_1,z_2)$ into the new coordinate system $(x,y)$ with the origin positioned as shown in Fig.~\ref{fig1}~(b). The tunnel exchange splitting integral in Eq.~(\ref{Egu}) now takes the form
\begin{equation}
J_{X^{\ast}}(\Delta Z)=\int_{\!-\Delta Z/\!\sqrt{2}}^{\Delta Z/\!\sqrt{2}}\!dy\left|\psi_{X^\ast}(x,y)\frac{\partial\psi_{X^\ast}(x,y)}{\partial x}\right|_{x=0},
\label{JXast}
\end{equation}
where $\psi_{X^\ast}(x,y)$ is the ground-state wave function of the Schr\"{o}dinger equation with the Hamiltonian (\ref{biexcham}) modified to the negative trion case, as discussed above, and then transformed to the $(x,y)$ coordinates. The tunnel exchange current integral $J_{X^{\ast}}(\Delta Z)$ is due to the electron position exchange relative to the hole (see Fig.~\ref{fig2}). This corresponds to the tunneling of the entire three particle system between the two equivalent indistinguishable configurations of the two excitons sharing the same hole in the configuration space $(z_1,z_2)$, given by the pair of minima at $z_1\!=\!z_2\!=\!0$ and $z_1\!=\!-z_2\!=\!\Delta Z$ in Fig.~\ref{fig1}~(b). Such a tunnel exchange interaction is responsible for the coupling of the three particle system to form a stable trion state.

Like in the case of the biexciton, one seeks the function $\psi_{X^{\ast}}(x,y)$ in the form
\begin{equation}
\psi_{X^{\ast}}(x,y)=\psi_0(x,y)\exp[-S_{X^{\ast}}(x,y)]\,,
\label{psixy}
\end{equation}
with $\psi_0(x,y)$ given by Eq.~(\ref{psi0xy}), where $S_{X^{\ast}}(x,y)$ is assumed to be a \emph{slowly} varying function to take into account the deviation of $\psi_{X^{\ast}}$ from $\psi_0$ in the "tail area" of $\psi_0$ due to the tunnel exchange coupling to another equivalent isolated two-exciton state centered at $z_1\!=\Delta Z$, $z_2\!=\!-\Delta Z$ (or $x\!=\!\Delta Z/\sqrt{2}$, $y\!=\!0$). Substituting Eq.~(\ref{psixy}) into the Schr\"{o}dinger equation with the negative trion Hamiltonian pre-transformed to the $(x,y)$ coordinates, one obtains in the region of interest
\[
\frac{\partial S_{X^{\ast}}}{\partial x}=\frac{\nu_0}{\Delta Z/\sqrt{2}-x}
\]
($|x|\!<\!\Delta Z/\sqrt{2}$, cut-off $z_0$ dropped) up to terms of the order of the second derivatives of~$S_{X^{\ast}}$. This is to be solved with the boundary condition $S_{X^{\ast}}(-\Delta Z/\sqrt{2},y)\!=\!0$ coming from the requirement $\psi_{X^{\ast}}(-\Delta Z/\sqrt{2},y)\!=\!\psi_0(-\Delta Z/\sqrt{2},y)$, to result in
\begin{equation}
S_{X^{\ast}}(x,y)=\nu_0\ln\frac{\sqrt{2}\Delta Z}{\Delta Z/\sqrt{2}-x}.
\label{sxy}
\end{equation}

After plugging Eqs.~(\ref{sxy}) and (\ref{psixy}) into Eq.~(\ref{JXast}), and retaining only the leading term of the integral series expansion in powers of $\nu_0$ subject to $\Delta Z>1$, one obtains
\begin{equation}
J_{X^{\ast}}(\Delta Z)=\frac{2}{2^{2\nu_0}\nu_0^3}\Delta Z\,e^{-2\Delta Z/\nu_0}.
\label{JXastfin}
\end{equation}
Inserting this into the right-hand side of Eq.~(\ref{Egu}), one sees that the ground state energy $E_g$ of the three particle system goes through the negative minimum (the trion state) as $\Delta Z$ increases. The minimum occurs at $\Delta Z_0=\nu_0/2$, whereby the trion binding energy in atomic units takes the form
\[
E_{X^{\!\ast}}=-J_{X^{\ast}}(\nu_0/2)=-\frac{1}{e\,2^{2\nu_0}\nu_0^2}.
\]
In absolute units, expressing $\nu_0$ in terms of $E_X$, one obtains
\begin{equation}
E_{X^{\!\ast}}=-\frac{|E_X|}{e\,2^{2\sqrt{Ry^\ast/|E_X|}}}.
\label{Exstar}
\end{equation}


\begin{figure}[t]
\epsfxsize=12.0cm\centering{\epsfbox{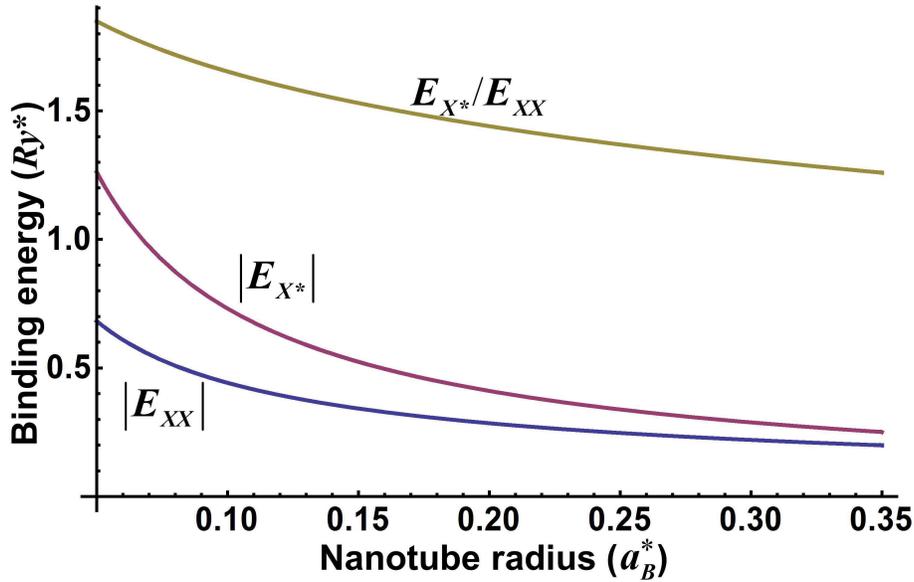}}\caption{Trion binding energy, biexciton binding energy and their ratio given by Eqs.~(\ref{Exstar}), (\ref{Exx}) and (\ref{ExstarExx}), respectively, with $|E_X|\!=\!Ry^\ast\!/r^{0.6}$ as functions of the dimensionless nanotube radius.}\label{fig3}
\end{figure}

\section{Comparative analysis of $E_{X\!X}$ and $E_{X^{\!\ast}}$ in quasi-1D}

From Eqs.~(\ref{Exx}) and (\ref{Exstar}), one has the trion-to-biexciton binding energy ratio as follows
\begin{equation}
\frac{E_{X^{\!\ast}}}{E_{X\!X}}=3\!\left(\frac{3}{2e}\right)^{\!2\sqrt{Ry^\ast/|E_X|}}\!\!.
\label{ExstarExx}
\end{equation}
If one now assumes $|E_X|\!=\!Ry^\ast\!/r^{0.6}$ ($r$ is the dimensionless CN radius, or transverse confinement size for quasi-1D nanostructure in general) as was demonstrated earlier by variational calculations\cite{Pedersen03} to be consistent with many quasi-1D models,\cite{Bondarev09PRB,Ogawa91,Loudon} then one obtains the $r$-dependences of $|E_{X^{\!\ast}}|$, $|E_{X\!X}|$ and $E_{X^{\!\ast}}\!/E_{X\!X}$ shown in Fig.~\ref{fig3}. The trion and biexciton binding energies both decrease with increasing $r$ --- in such a way that their ratio remains greater than unity for small enough $r$ --- in full agreement with the experiments by Colombier et al.\cite{Colombier12} and Yuma et al.\cite{Yuma13} However, since the factor $3/2e$ in Eq.~(\ref{ExstarExx}) is less than one, the ratio can also be less than unity for $r$ large enough (but not too large, so that our configuration space method still works).

As $r$ goes down, on the other hand, the biexciton-to-exciton binding energy ratio $|E_{X\!X}/E_X|$ in Eq.~(\ref{Exx}) slowly grows, approaching the pure 1D limit $1/3e\approx0.12$. Similar tendency can also be traced in the Monte-Carlo simulation data of Ref.\cite{Kammerlander07} The equilibrium inter-exciton center-of-mass distance in the biexciton complex goes down with decreasing $r$ as well, $\Delta Z_0=\nu_0/2=\!1/(2\sqrt{|E_X|}\,)$ (atomic units). This supports experimental evidence for enhanced exciton-exciton annihilation in small diameter CNs.\cite{THeinz,Valkunas,Kono} The trion-to-exciton binding energy ratio $|E_{X^\ast}/E_X|$ of Eq.~(\ref{Exstar}) increases with decreasing $r$ faster than $|E_{X\!X}/E_X|$ (Fig.~\ref{fig3}), to yield $E_{X^\ast}/E_{X\!X}\!\approx\!3$ as the pure 1D limit for the trion-to-biexciton binding energy ratio.


\begin{figure}[t]
\epsfxsize=11.5cm\centering{\epsfbox{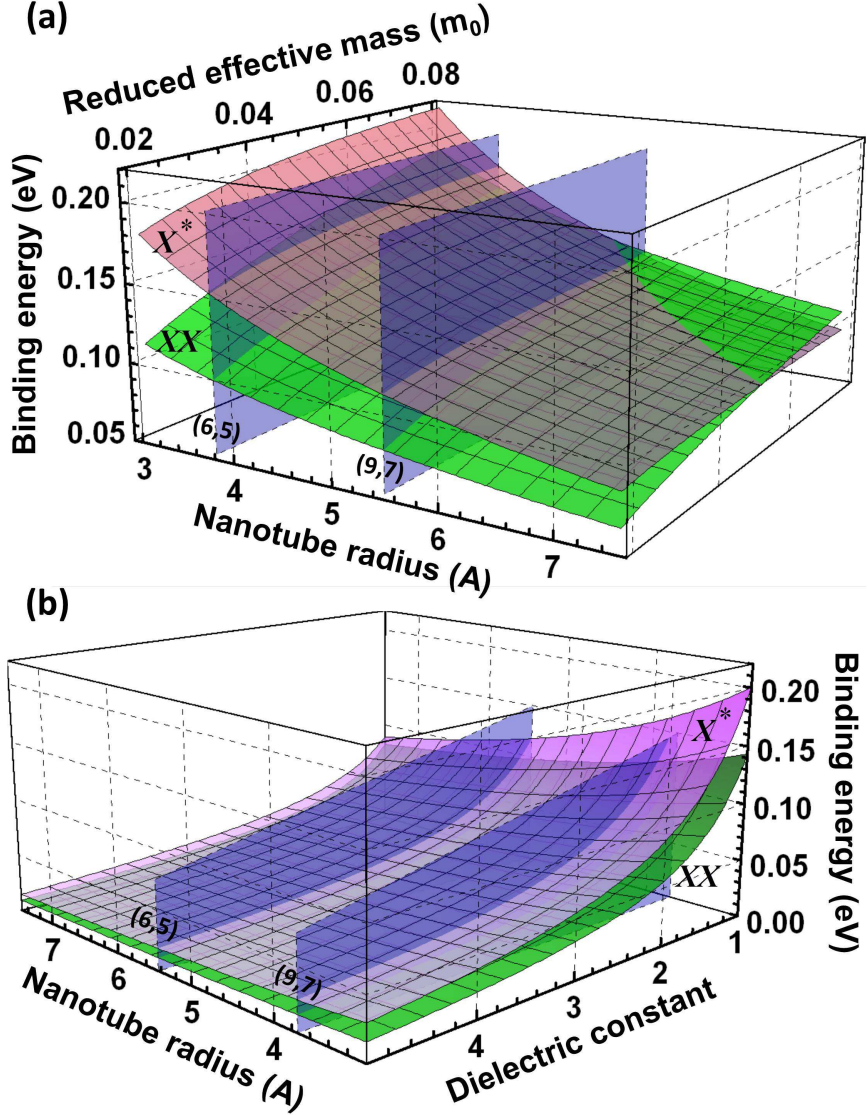}}\caption{Trion ($X^\ast$) and biexciton ($X\!X$) binding energies given by Eqs.~(\ref{Exstar}) and (\ref{Exx}) with $|E_X|\!=\!Ry^\ast\!/r^{0.6}$, as functions of the CN radius and $\mu$ with $\varepsilon\!=\!1$ (a), and as functions of the CN radius and $\varepsilon$ with $\mu\!=\!0.04$ (b). Vertical parallel planes indicate the radii of the (6,5) and (9,7) CNs studied experimentally.}\label{fig4}
\end{figure}

When $E_{X^{\!\ast}}\!/E_{X\!X}$ is known, one can use Eq.~(\ref{ExstarExx}) to estimate the effective Bohr radii $a^\ast_B$ for the excitons in the CNs of known radii. For example, substituting $1.46$ for the 0.75~nm diameter (6,5) CN and $1.42$ for the 1.09~nm diameter (9,7) CN, as reported by Yuma et al.\cite{Yuma13} and Colombier et al.,\cite{Colombier12} respectively, into the left hand side of the transcendental equation~(\ref{ExstarExx}) and solving it for $a^\ast_B$, one obtains the effective exciton Bohr radius $a^\ast_B\!=\!2$~nm and $2.5$~nm for the (6,5) CN and (9,7) CN, respectively. This agrees reasonably with previous estimates.\cite{Pedersen03,Yuma13}

In general, the binding energies in Eqs.~(\ref{Exstar}) and (\ref{Exx}) are functions of the CN radius (transverse confinement size for a quasi-1D semiconductor nanowire), $\mu$ and~$\varepsilon$. Figures~\ref{fig4}~(a) and~\ref{fig4}~(b) show their 3D plots at fixed $\varepsilon\;(=\!1)$ and at fixed $\mu\;(=\!0.04)$, respectively, as functions of two remaining variables. The reduced effective mass $\mu$ chosen is typical of large radius excitons in small-diameter CNs.\cite{Jorio05} The unit dielectric constant $\varepsilon$ assumes the CN placed in air and the fact that there is no screening in quasi-1D semiconductor systems both at short and at large electron-hole separations.\cite{Louie09} This latter assumption of the unit background dielectric constant remains legitimate for \emph{small} diameter ($\lesssim\!1$~nm) semiconducting CNs in dielectric screening environment, too, --- for the lowest excitation energy exciton in its ground state of interest here (not for its excited states though), in which case the environment screening effect is shown by Ando to be negligible,\cite{Ando2010} diminishing quickly with the increase of the effective distance between the CN and dielectric medium relative to the CN diameter.

Figure~\ref{fig4}~(a) can be used to evaluate the relative stability of the trion and biexciton complexes in quasi-1D semiconductors. One sees that whether the trion or the biexciton is more stable (has the greater binding energy) in a particular quasi-1D system depends on $\mu$ and on the characteristic transverse size of the nanostructure. In strongly confined quasi-1D systems with relatively small $\mu$, such as small-diameter CNs, the trion is generally more stable than the biexciton. In less confined quasi-1D structures with greater $\mu$ typical of semiconductors,\cite{Cardona} the biexciton is more stable than the trion. This is a generic peculiarity in the sense that it comes from the tunnel exchange in the quasi-1D electron-hole system in the configuration space. Greater $\mu$, while not affecting significantly the single charge tunnel exchange in the trion complex, makes the neutral biexciton complex generally more compact, facilitating the mixed charge tunnel exchange in it and thus increasing the stability of the complex. From Fig.~\ref{fig4}~(b) one sees that this generic feature is not affected by the variation of $\varepsilon$, although the increase of $\varepsilon$ decreases the binding energies of both excitonic complexes --- in agreement both with theoretical studies\cite{Ronnow10} and with experimental observations of lower binding energies (compared to those in CNs) of these complexes in conventional semiconductor nanowires.\cite{Forchel,Crottini,Sidor,Gonzales,Schuetz} The latter are self-assembled nano\-structures of one (transversely confined) semiconductor embedded in another (bulk) semiconductor with the characteristic transverse confinement size typically greater than that of small diameter CNs, and so both inside and outside material dielectric properties matter there.


\begin{figure}[t]
\epsfxsize=12.0cm\centering{\epsfbox{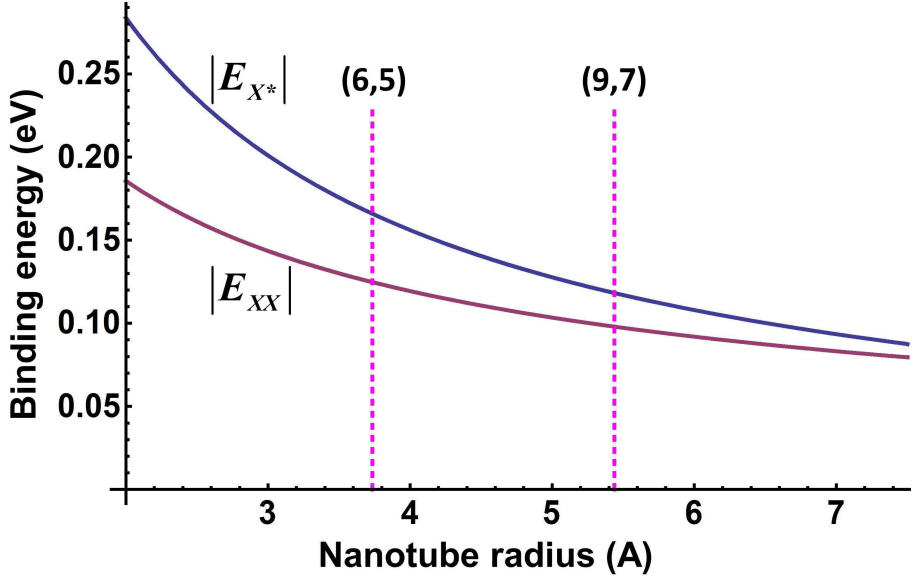}}\caption{Cross-section of Fig.~\ref{fig4}(a) at $\mu\!=\!0.04$ showing the relative behavior of the trion and biexciton binding energies in semiconducting CNs of increasing radius.}\label{fig5}
\end{figure}

Figure~\ref{fig5} shows the cross-section of Fig.~\ref{fig4}~(a) taken at $\mu\!=\!0.04$ to present the relative behavior of $|E_{X^{\!\ast}}|$ and $|E_{X\!X}|$ in semiconducting CNs of increasing radius. Both $|E_{X^{\!\ast}}|$ and $|E_{X\!X}|$ decrease, and so does their ratio, as the CN radius increases. From the graph, $|E_{X^{\!\ast}}|\!\approx\!170$ and $125$~meV, $|E_{X\!X}|\!\approx\!120$ and $95$~meV, for the (6,5) and (9,7) CNs, respectively. This is to be compared with $190$ and $130$~meV for the (6,5) CN versus $150$ and $106$~meV for the (9,7) CN reported experimentally.\cite{Colombier12,Yuma13} One sees that, as opposed to perturbative theories,\cite{Watanabe12} the present configuration space theory underestimates experimental data just slightly, most likely due to the standard variational treatment limitations. It does explain well the trends observed, and so the graph in Fig.~\ref{fig5} can be used as a guide for trion and biexciton binding energy estimates in small diameter ($\lesssim\!1$~nm) nanotubes.

\section{Configuration space method as applied to quasi-2D systems}

\begin{figure}[t]
\epsfxsize=12.0cm\centering{\epsfbox{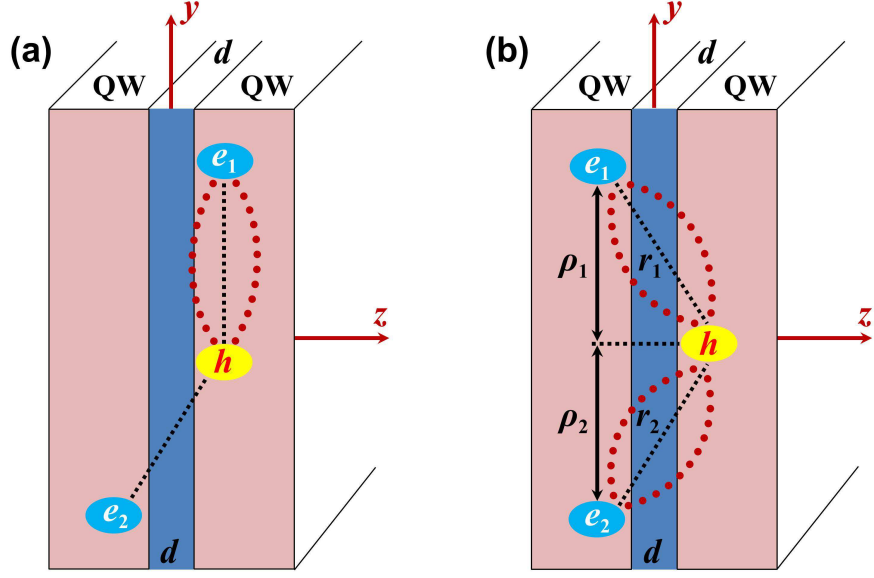}}\vskip-0.5cm\caption{Schematic view of the trion formed by an electron and a direct exciton (a) and that formed by an electron and an indirect exciton (b) in the coupled quantum well nanostructure. In (a), exciton configurations $(h-e_1)$ and $(h-e_2)$ are inequivalent. In (b), they are equivalent and so the configuration space method can be used to evaluate the binding energy of the trion complex.}\label{fig6}
\end{figure}

Recently, there has been a considerable interest in studies of optical properties of coupled quantum wells (CQWs).\cite{Kezer14,Kezer14jmpb,Govorov13,Muljarov12,Jeremy11,Gossard11,Govorov11,Butov09,Butov09prb,Snoke09,Bloch06,Snoke06} The CQW semiconductor nanostructure (Fig.~\ref{fig6}) consists of two identical semiconductor quantum wells separated by a thin barrier layer of another semiconductor. The tunneling of carriers through the barrier makes two wells electronically coupled to each other. As a result, an electron (a hole) can either reside in one of the wells, or its wave function can be distributed between both wells. A Coulomb bound electron-hole pair residing in the same well forms a direct exciton [Fig.~\ref{fig6}~(a)]. If the electron and hole of a pair are located in different wells, then an indirect exciton is formed [Fig.~\ref{fig6}~(b)].


Physical properties of CQWs can be controlled by using external electro- and magnetostatic fields. (See, e.g., Refs.\cite{Govorov13,Muljarov12} and refs. therein.) For example, applying the electrostatic field perpendicular to the layers increases the exciton radiative lifetime due to a reduction in the spatial overlap (contact density) between the electron and hole wave functions. As this takes place, the exciton binding energy reduces due to an increased electron-hole separation to make the exciton less stable against ionization, in contrast with the exciton magnetostatic stabilization effect under the same geometry.\cite{Govorov13} The tunneling effect is also enhanced as the electric field allows the carriers to leak out of the system, resulting in a considerable shortening of the photoluminescence decay time.

CQWs embedded into Bragg-mirror microcavities show a special type of voltage-tuned exciton polaritons, which can be used for low-threshold power polariton lasing.\cite{Jeremy11} New non-linear phenomena are also reported for these CQW systems both theoretically and experimentally, such as Bose condensation\cite{Kezer14} and parametric oscillations\cite{Bloch06} of exciton polaritons. For laterally confined CQW structures, experimental evidence for controllable formation of multiexciton Wigner-like molecular complexes of indirect excitons (single exciton, biexciton, triexciton, etc.) was reported recently.\cite{Govorov13} Trion complexes formed both by direct and by indirect excitons, as sketched in Figs.~\ref{fig6}~(a) and (b), were observed in CQWs as well.\cite{Shields97} All these findings make CQWs a much richer system capable of new developments in fundamental quantum physics and nanotechnology as compared to single quantum wells.\cite{Jeremy11,Gossard11,Govorov11,Butov09,Bloch06} They open up new routes for non-linear coherent optical control and spinoptronics applications with quasi-2D semiconductor CQW nanostructures.

Very recently, the problem of the trion complex formation in CQWs was studied theoretically in great detail for trions composed of a \emph{direct} exciton and an electron (or a hole) located in the neighboring quantum well\cite{Kezer14jmpb} [as sketched in Fig.~\ref{fig6}~(a)]. Significant binding energies are predicted on the order of $10$~meV at interwell separations $d\sim\!10-20$~nm for the lowest energy positive and negative trion states, to allow one suggest a possibility for trion Wigner crystallization. Figure~\ref{fig6}~(b) shows another possible trion complex that can also be realized in CQWs. Here, the trion is composed of an \emph{indirect} exciton and an electron (or a hole) in such a manner as to keep two same-sign particles in the same quantum well with the opposite-sign particle being located in the neighboring well. This can be viewed as the two equivalent configurations of the three-particle system in the configuration space ($\rho_1$, $\rho_2$) of the two \emph{independent} in-plane projections of the relative electron-hole distances $r_1$ and $r_2$ in the two \emph{equivalent} indirect excitons sharing the same electron, or the same hole as shown in Fig.~\ref{fig6}~(b). Such a three-particle system in the quasi-2D semiconductor CQW nanostructure is quite analogous to the quasi-1D trion presented in Sec.~3 above [cf. Fig.~\ref{fig6}~(b) and Fig.~\ref{fig2}]. Therefore, the Landau-Herring configuration space approach can be used here as well to evaluate the binding energy for this special case of the quasi-2D trion state.

Following is a brief outline of how one could proceed with the configuration space method to obtain the ground state binding energy for the quasi-2D trion complex sketched in Fig~\ref{fig6}~(b). A complete analysis of the problem will be presented elsewhere. The method requires knowledge of the ground state characteristics of the indirect exciton (abbreviated as "$I\!X$" in what follows). Specifically, one needs to know the quasi-2D ground state energy $E_{I\!X}(d)$ and corresponding \emph{in-plane} relative electron-hole motion wave function $\psi_{I\!X}(\rho,d)$ for the indirect exciton in the CQW system with the interwell distance $d$. These can be found by solving the radial Scr\"{o}dinger equation that is obtained by decoupling radial relative electron-hole motion in the cylindrical coordinate system with the $z$-axis being perpendicular to the QW layers [see Fig~\ref{fig6}~(b)]. Such an equation was derived and analyzed previously by Leavitt and Little.\cite{Leavitt90} The energy and the wave function of interest are as follows
\begin{eqnarray}
E_{I\!X}(d)=\lambda^2-\frac{4\lambda+4\lambda^4d^2E_1(\lambda d)\exp(2\lambda d)}{1+2\lambda d}\,,\nonumber\\[-0.2cm]
\label{indirect}\\[-0.2cm]
\psi_{I\!X}(\rho,d)=N\exp[-\lambda(\sqrt{\rho^2+d^2}-d)]\,,\hskip0.5cm\nonumber
\end{eqnarray}
where $E_1(x)\!=\!\int_{x}^{\infty}(e^{-t}/t)\,dt$ is the exponential integral, $\lambda=\!\lambda(d)\!=\!2/(1+2\sqrt{d}\,)$, the normalization constant $N$ is determined from the condition
\[
\int_{0}^{\infty}\!|\psi_{I\!X}(\rho,d)|^2\rho\,d\rho=1,
\]
and all quantities are measured in atomic units as defined in Sec.~2.


\begin{figure}[t]
\epsfxsize=12.0cm\centering{\epsfbox{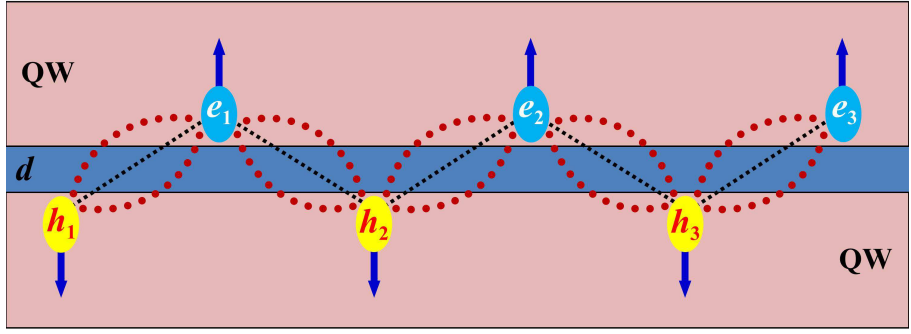}}\caption{Schematic of the charge-neutral spin-aligned Wigner crystal structure formed by two trions. They are one composed of the indirect exciton $(h_1-e_1)$ and the hole $h_2$, and another one composed of the indirect exciton $(h_3-e_3)$ and the electron $e_2$ [cf. Fig.~\ref{fig6}~(b)]. The structure can also be viewed as a triexciton, a coupled state of three indirect singlet excitons.}\label{fig7}
\end{figure}

With Eq.~(\ref{indirect}) in place, one can work out strategies for tunneling current calculations in the configuration space ($\rho_1$, $\rho_2$) of the two \emph{independent} in-plane projections of the relative electron-hole distances $r_1$ and $r_2$ in the two \emph{equivalent} indirect excitons as shown in Fig.~\ref{fig6}~(b). Both tunneling current responsible for the trion complex formation and that responsible for the biexciton complex formation can be obtained in full analogy with how it was done above for the respective quasi-1D complexes. The only formal difference now is the change in the phase integration volume from $dz_1dz_2$ to $\rho_1d\rho_1\rho_2d\rho_2$. Minimizing the tunneling current with respect to the center-of-mass-to-center-of-mass distance of the two equivalent indirect excitons results in the binding energy of a few-particle complex of interest. Note that the method applies to the complexes formed by \emph{indirect} excitons only as they allow equivalent configurations for a few-particle system to tunnel throughout in the configuration space ($\rho_1$, $\rho_2$), thereby forming a respective (tunnel) coupled few-particle complex.

Binding energy calculations for indirect exciton complexes in semiconductor CQW nanostructures are important to understand the principles of the more complicated electron-hole structure formation such as that shown in Fig.~\ref{fig7}. This is a coupled charge-neutral spin-aligned Wigner-like structure formed by two trions, one positively charged and another one negatively charged. The entire structure is electrically neutral, and it has an interesting electron-hole spin alignment pattern. This structure can also be viewed as a triexciton, a coupled state of three indirect singlet excitons. One could also imagine a Wigner-like crystal structure formed by unequal number of electrons and holes, as opposed to that in Fig.~\ref{fig7}, whereby the entire coupled structure could possess net charge and spin at the same time to allow precise electro- and magnetostatic control and manipulation by its optical and spin properties. Such Wigner-like electron-hole crystal structures in CQWs might be of great interest for spinoptronics applications.

All in all, indirect excitons, biexcitons and trions formed by indirect excitons are those building blocks that control the formation of more complicated Wigner-like electron-hole crystal structures in CQWs. The configuration space method presented here allows one to study the binding energies for these building blocks as functions of CQW system parameters, and thus to understand how stable electron-hole Wigner crystallization could possibly be in these quasi-2D nanostructures. The method should also work well for biexciton and trion complexes in quasi-2D self-assembled transition metal dichalcogenide heterostructures, where electrons and holes accumulated in the opposite neighboring monolayers are recently reported to form indirect excitons with new exciting properties such as increased recombination time\cite{Ceballos14} and vanishing high-temperature viscosity.\cite{Fogler14}

\section{Conclusion}

Presented herein is a universal configuration space method for binding energy calculations of the lowest energy neutral (biexciton) and charged (trion) exciton complexes in spatially confined quasi-1D semiconductor nanostructures. The method works in the effective two-dimensional configuration space of the two relative electron-hole motion coordinates of the two non-interacting quasi-1D excitons. The biexciton or trion bound state forms due to under-barrier tunneling between equivalent configurations of the electron-hole system in the configuration space. Tunneling rate controls the binding strength and can be turned into the binding energy by means of an appropriate variational procedure. Quite generally, trions are shown to be more stable (have greater binding energy) than biexcitons in strongly confined quasi-1D structures with small reduced electron-hole masses. Biexcitons are more stable than trions in less confined structures with large reduced electron-hole masses. A universal crossover behavior is predicted whereby trions become less stable than biexcitons as the transverse size of the quasi-1D nanostructure increases.

An outline is given of how the method can be used for electron-hole complexes of indirect excitons in quasi-2D semiconductor systems such as coupled quantum wells and van der Waals bound transition metal dichalcogenide heterostructures. Here, indirect excitons, biexcitons, and trions formed by indirect excitons control the formation of more complicated Wigner-like electron-hole crystal structures. The configuration space method can help develop understanding of how stable Wigner crystallization could be in these quasi-2D nanostructures. Wigner-like electron-hole crystal structures are of great interest for future spinoptronics applications.

\section*{Acknowledgments}

This work is supported by the US Department of Energy (DE-SC0007117). Discussions with David Tomanek (Michigan State U.), Roman Kezerashvili (NY$\,\!$CityTech), and Masha Vladimirova (U. Montpellier, France) are acknowledged. I.V.B. thanks Tony Heinz (Stanford U.) for pointing out Ref.\cite{Louie09} of relevance to this work.

\end{document}